\documentclass[twocolumn,colorlinks]{goose-article}

\usepackage{enumitem}

\usepackage{booktabs}
\usepackage[toc,page,titletoc]{appendix}
\crefname{supp}{Supplement}{Supplements}

\hypersetup{pdfauthor={T.W.J. de Geus}}

\title{Short-range depinning in the presence of velocity-weakening}

\author[1]{Tom~W.J.~de~Geus}
\author[1,2]{Matthieu~Wyart}
\affil[1]{Institute of Physics, \'{E}cole Polytechnique F\'{e}d\'{e}rale de Lausanne (EPFL), Switzerland}
\affil[2]{Department of Physics and Astronomy, Johns Hopkins University, Baltimore, MD, USA}

\newcommand{\detail}[1]{}

\begin{document}

\twocolumn[
    \begin{@twocolumnfalse}

        \maketitle

        \begin{abstract}
            \noindent
            Phenomena including friction and earthquakes are complicated by the joint presence of disorder and non-linear instabilities, such as those triggered by the presence of velocity weakening.
            In \cite{deGeus2022}, we provided a theory for the nucleation of flow and the magnitude of hysteresis, building on recent results on disorder-free systems described by so called rate-and-state descriptions of the frictional interface, and treating disorder perturbatively.
            This theory was tested for models of frictional interfaces, where long range elastic interactions are present.
            Here we test it for short-range depinning, and confirm that (i) nucleation is triggered by avalanches, governed by a critical point at some threshold force $f_c$ close to the minimum of the flow curve and that (ii) due to an armouring mechanism by which the elastic manifold displays very little plasticity after a big slip event, very slowly decaying finite size effects dominate the hysteresis magnitude, with an exponent we can relate to other observables.

        \end{abstract}

    \end{@twocolumnfalse}
]

\section{Introduction}

Elastic interfaces pinned by disorder are found in various systems, including crack fronts in fracture \cite{Bouchaud1997}, frictional interfaces \cite{Scholz1998}, domain walls in magnets \cite{Dahmen1996, Zapperi1998}, sliding charge density waves \cite{Narayan1992} or vortex lattices in superconductors \cite{Blatter1994}.
In the absence of temperature, the interface is pinned by impurities, but undergoes a \emph{depinning transition} \cite{Kardar1998,Fisher1998,Narayan1993} at some critical force (or stress, or magnetic field).
When inertia is not present, the physics is well-understood: the interface moves via large reorganizations called avalanches, and the velocity of the interface grows above the depinning threshold where it vanishes with some critical exponent.

\begin{figure}[htp]
    \subfloat{\label{fig:nonmonotic:a}}
    \subfloat{\label{fig:nonmonotic:b}}
    \centering
    \includegraphics[width=\linewidth]{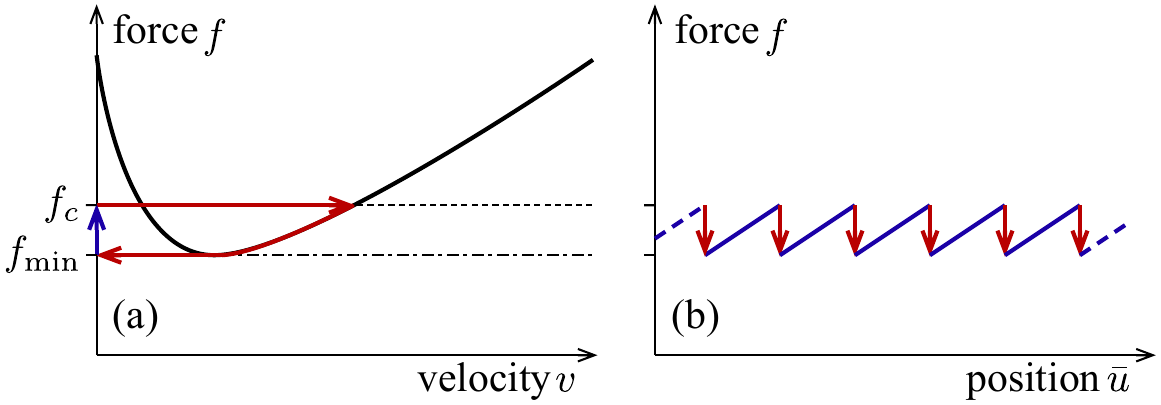}
    \caption{
        [Sketch] \protect\subref{fig:nonmonotic:a} When the flow curve (force $f$ \emph{vs} velocity $v$) is non-monotonic, with a minimum at $f = f_{\min}$, there exists a threshold force $f_c$ beyond which the static phase is unstable to system-spanning events.
        \protect\subref{fig:nonmonotic:b}
        A finite hysteresis is predicted in the thermodynamic limit, such that a system driven quasi-statically through a weak spring displays stick-slip (by ``a weak spring'' we mean that $k_f$ of Eq.~\eqref{eq:motion} is small: $k_f \leq \mathcal{O}(1 / L^2$) with $L$ the system size, see text in Sec.~\ref{sec:model}).
        Thereby, power law distributed avalanches protect the interface from building up a load $f > f_c$ where it is unstable.
        After the instability, the interface unloads to $f = f_{\min}$ while slipping (see corresponding cycle in panel \protect\subref*{fig:nonmonotic:a} with the same colour coding).
    }
    \label{fig:nonmonotic}
\end{figure}

In other systems, such as frictional interface \cite{Baumberger2006,Tromborg2015a}, crack fronts in brittle materials \cite{Ramanathan1997} or wetting on rough solid surfaces \cite{Ertas1992,Prevost2002}, a sufficient inertia or other effects can induce the flow curve to be non-monotonic \cite{Marchetti2005a,Nicolas2016,Karimi2016}.
Indeed, in the presence of inertial effects, flow generates acoustic waves, that can make it easier to jump over local barriers induced by disorder, thus generating more acoustic waves, in turn accelerating flow.
Such positive feedback, or `inertial heating', can lead to velocity-weakening, causing instabilities and stick-slip.
The flow curve of such an inertial system is non-monotonic, as sketched in \cref{fig:nonmonotic:a}, whereby, at low velocities, the flow resistance decreases with increasing velocity due the described `inertial heating', while at high velocities, the flow resistance increases with velocity as viscous dissipation dominates the `inertial heating'.
How instabilities are triggered as the force is slowly increased is a central question to various fields, including earthquake science and tribology.
Understanding it remains a theoretical challenge, as one needs to deal both with disorder and the presence of non-linear instabilities.
This question relates to what controls the magnitude of hysteresis effects apparent when comparing the starting point of the flow upon increasing the applied load to its stopping point upon decreasing it again.

\begin{figure}[htp]
    \subfloat{\label{fig:pre:flow}}
    \subfloat{\label{fig:pre:stick-slip}}
    \subfloat{\label{fig:pre:bimodal}}
    \subfloat{\label{fig:pre:ellc}}
    \centering
    \includegraphics[width=\linewidth]{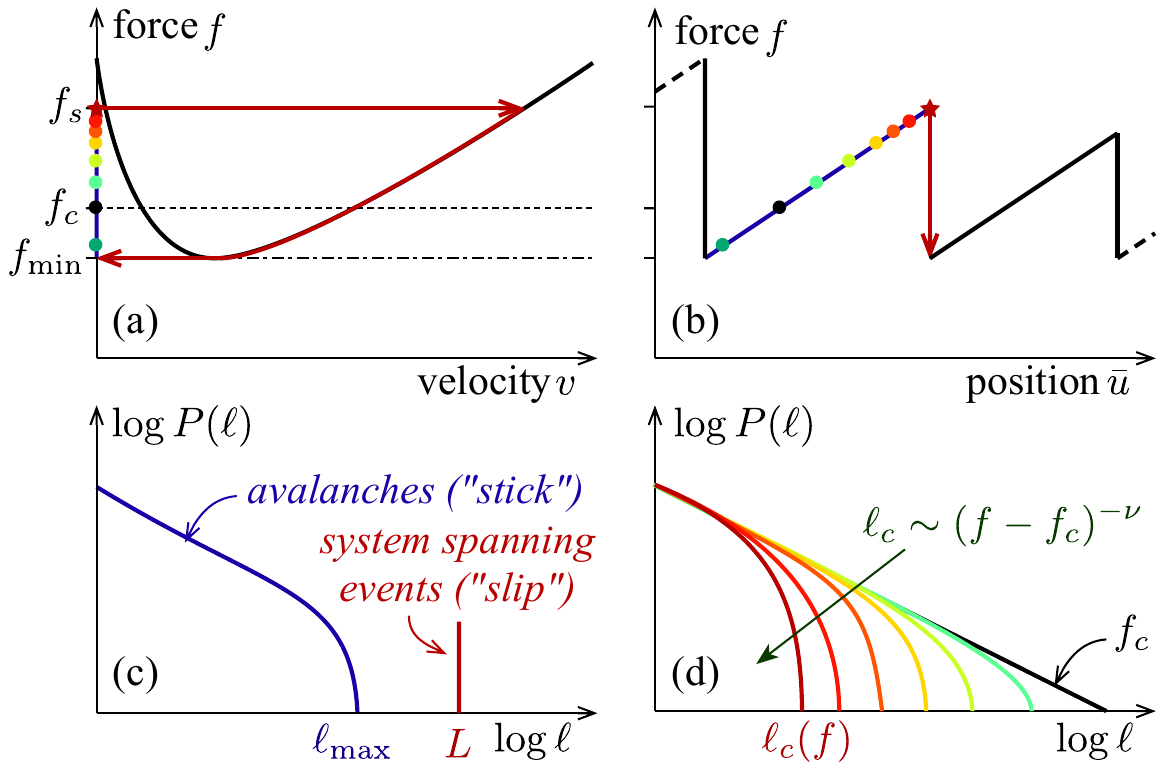}
    \caption{
        [Sketch] (\protect\subref*{fig:pre:flow}-\protect\subref*{fig:pre:stick-slip}) In a quasi-statically loaded finite system with disorder, avalanches occur as the force increases.
        They nucleate slip once their extension $\ell > \ell_c \sim (f - f_c)^{-\nu}$ if $f > f_c$.
        The force at which nucleation occurs fluctuates, we denote by $f_s$ its means (and study below how the magnitude of hysteresis $\Delta f=f_s-f_c$ depends on $L$).
        \protect\subref{fig:pre:bimodal} The corresponding distribution of event extensions, $P(\ell)$, during a quasi-static cycle is bimodal, with avalanches up to a scale $\ell_{\max}(f_s)$, and system-spanning events ($\ell = L$).
        See \cref{sec:criticality} for numerical evidence of bimodality.
        \protect\subref{fig:pre:ellc} To study the properties of avalanches in an infinite system, and to quantify $\ell_c$, we trigger avalanches at different forces $f$.
        The distribution of their linear extension $\ell$ is scale free at $f_c$ while at $f > f_c$ avalanches transition to system-spanning events if $\ell > \ell_c$ (system-spanning events are excluded from the shown distributions).
    }
    \label{fig:pre}
\end{figure}

At least three scenarios have been proposed for how the depinning transition is affected by inertia:

\begin{enumerate}[wide, labelwidth=!, labelindent=0pt, label={\emph{Scenario \Alph*}:}, ref={scenario \Alph*}]

    \item \label{item:scenario:a}
    The depinning transition becomes first order as soon as inertia is present.
    For a finite amount of inertia, motion is triggered by small avalanches, and the hysteresis loop is of finite magnitude in the thermodynamic limit.
    This picture was obtained by treating inertia as a perturbation around the usual depinning behaviour \cite{Fisher1997}.
    It does not occur in mean-field \cite{Marchetti2005a} nor in more recent numerical observations \cite{Nicolas2016,Karimi2016,Salerno2012,Salerno2013}, where for mild inertial effects evidence of a continuous transition exists.

    \item \label{item:scenario:b}
    For small inertia, the flow curve is still monotonic with some modified threshold force $f_c^\downarrow$ \cite{Schwarz2003}.
    Flow is critical at $f_c^\downarrow$, and in the same universality class as when inertia is absent.
    The hysteresis loop vanishes in the thermodynamic limit, but very slowly, due to a combination of two effects.
    First, nucleation of flow emerges from rare avalanches exceeding a size $\ell_c\sim(f-f_c^\downarrow)^{-\tilde \nu}$, where $\tilde \nu$ is a critical exponent characterizing the depinning transition in the absence of inertia.
    Second, only rare sites can yield under loading after a slip event and trigger an avalanche.
    Limitations of this approach are the following.
    \emph{Limitation B1:} It was supported by an automaton model where inertia is modelled as a brief overshoot of the force change in sites surrounding rearrangements.
    In \cite{Maimon2004}, it was argued that for more generic automaton models, a finite hysteresis should be present, while flow is still governed by a continuous transition at $f_c^\downarrow$.
    These results indicate the need to go beyond automaton models and model inertia more accurately.
    \emph{Limitation B2:} This approach focused on small inertial effects, and did not capture macroscopic velocity weakening.
    \emph{Limitation B3:} The nucleation picture could not be tested precisely because in systems of limited size, flow is triggered far from $f_c^\downarrow$.

    \item \label{item:scenario:c}
    In a recent work \cite{deGeus2022}, we addressed these limitations and focused on the case where velocity weakening is present (thus \ref{item:scenario:c} needs not contradict \ref{item:scenario:b}, as these could each apply at different levels of inertial effect or damping coefficient).
    We built on the rate-and-state description of homogeneous frictional interfaces which includes from the start velocity weakening of the interface \cite{Dieterich1979,Rice1983}, and treat \emph{disorder} pertubatively.
    A result of the rate-and-state literature \cite{Zheng1998,Brener2018} central to our approach is pictured in \cref{fig:nonmonotic}: for a pure system (i.e.\ a system without disorder), there exists a threshold force $f_c$ (very) slightly above the minimum $f_{\min}$ of the flow curve\footnote{
        In Ref.~\cite{Brener2018}, $f_c$ is found by solving some coupled differential equations deriving from classical continuous Newtonian bulk dynamics coupled with a rate-and-state friction model of the interface.
        It is then observed that $f_c$ is just above the minimum of the flow curve, $f_{\min}$, for reasonable choices of parameters.
        We are not aware of a more profound explanation for this state of affair (which would be interesting to have)
    } $v(f)$, where a frictional interface can be destabilized and rupture -- a result that holds both for short and long-range elastic interactions \cite{Brener2018}.
    In \cite{deGeus2022}, we argued that in the presence of disorder, $f_c$ controls nucleation, triggered by avalanches whose extension goes beyond some length:
    \begin{equation}
        \label{eq:ellc}
        \ell_c(f)\sim(f-f_c)^{-\nu}.
    \end{equation}
    The exponent $\nu$ in the presence of velocity-weakening differs from $\tilde \nu$ observed in its absence.
    Nucleation is revealed by a bimodal distribution of event sizes \footnote{
        Such a bimodal distribution for earthquakes on a single fault was argued based on seismic data in \cite{Wesnousky1994}.
    }, corresponding of both avalanches and system-spanning events, as sketched \cref{fig:pre:bimodal}.
    Thus, nucleation from the static phase is controlled by a critical point at $f_c$, as pictured in \cref{fig:pre}.
    In this scenario, there theoretically exists a tiny finite hysteresis in the thermodynamic limit because $f_c > f_{\min}$, under the modelling assumptions of the rate-and-state framework~\cite{Brener2018}, as illustrated in \cref{fig:nonmonotic}.
    The flow regime, by contrast, displays a first order transition when the load is reduced.

    In this scenario, the hysteresis magnitude displays very slowly decaying finite size effects, for reasons similar to those put forward in \ref{item:scenario:b} and \cite{Schwarz2003} (except that theoretically the hysteresis magnitude does not decay to zero as $f_c > f_{\min}$~\cite{Brener2018}).
    In particular, after a large slip event, the system is depleted of weak regions about to trigger an avalanche (i.e.\ ``yield'').
    The density of such regions was argued to vanish at a small local yield threshold $x$ as a power law $P(x) \sim x^{\theta'}$ in \cite{deGeus2019,ElSergany2023}, an effect we called ``armouring'' of the interface.
    Such a power-law distribution is refereed to as pseudo-gap in the literature \cite{Muller2015,Lin2014}.
    It leads to a scaling law for the excess hysteresis magnitude as a function of system size.

    These results were tested in \cite{deGeus2019,deGeus2022} using a finite element model where inertia is treated by actually solving Newton's equation, and where disorder is modelled with a random potential acting on each element, inspired by a previous model by Jagla \cite{Jagla2007}.
    A central idea to test this picture of nucleation was to trigger avalanches at various force levels \footnote{
        More precisely, stress levels.
    } \cite{deGeus2019}, as the spontaneous triggering of avalanches under loading around $f_c$ is too rare to allow for statistical analysis.
    These studies focused on long-range interactions, where additional effects (such as ``radiation damping'') make this scenario slightly more complicated to test.

\end{enumerate}

In the present work, we test \ref{item:scenario:c} for a one dimensional elastic line pinned by disorder, with short-range elastic interactions.
We first verify our prediction on nucleation in \cref{eq:ellc}, which requires both to access the non-monotonic flow curve, used to extract $f_c$ (approximated as $f_{\min}$, from which our numerical measurements it cannot be distinguished, i.e.\ as far as our numerics can resolve $f_c \simeq f_{\min}$), as well as to study the statistics of avalanches artificially triggered at different forces along the stick-slip cycle.
Secondly, we confirm our predictions for how the magnitude of the hysteresis cycle depends on system size.
Thereto, we combine the avalanche and nucleation properties with the density of regions about to yield, $P(x)$, through the associated pseudo-gap exponent.

\section{Model}
\label{sec:model}

\paragraph{Dynamics}

We simulate a classical depinning model, however with inertia.
In particular, we consider a line in dimension $d = 1$ composed of $L$ particles that have a finite mass, interact elastically with their nearest neighbours, and experience a disordered pinning potential.
In our model, each particle (numbered $i = 1, 2, \ldots, L$) has a degree of freedom (the position $u_i$) that follows Newton dynamics.
The dynamics then read
\begin{equation}
    \label{eq:motion}
    m \ddot{u}_i =
    \underbrace{\vphantom{\big(}-\partial_u U_i}_{1) \equiv f_i^p} +
    \underbrace{\vphantom{\big(}\Delta u_i}_{2) \equiv f_i^n} +
    \underbrace{\vphantom{\big(}k_f (\bar{u} - u_i)}_{3) \equiv f_i^f} -
    \underbrace{\vphantom{\big(}\eta \dot{u}_i}_{4) \equiv f_i^d}
\end{equation}
(whereby we use a mass $m = 1$ that is uniform).
Here, $\dot{u}_i \equiv \partial_t u_i$ is the particle's velocity and $\ddot{u}_i \equiv \partial^2_t u_i$ is the particle's acceleration.
The restoring force on a particle $i$ is composed of the following terms.

1) Each particle is driven through a potential energy landscape $U_i$ that is piecewise linear elastic with uniform elastic constant $\mu = 1$ and random local barriers, see below.

2) Neighbouring particles interact elastically such that schematically $f_i^n = \Delta u_i$, the Laplacian of $u$.
We use the fourth order expansion \cite{Rosso2001,Rosso2002a} such that in $d = 1$ dimensions, $f_i^n = k_2 \big[ u_{i - 1} - 2 u_i + u_{i + 1} \big] + k_4 \big[ (u_{i + 1} - u_i)^3 + (u_i - u_{i - 1})^3 \big]$ with $k_2 = k_4 = 1$ (see \cref{sec:interactions} for a derivation), so as to avoid an unrealistic roughness exponent $\zeta > 1$ known to occur with overdamped dynamics.
Moreover, we assume periodic boundary conditions $u_{L + i} = u_i$ for any $i$ (in particular $u_1 = u_L$).
Henceforth, we express length in units of the discretization, such that the system's extension is $L$.

3) We drive by connecting each particle with a weak spring of stiffness $k_f$ to a ``driving frame'' whose position $\bar{u}$ is prescribed, see below.

\begin{figure}[htp]
    \centering
    \includegraphics[width=\linewidth]{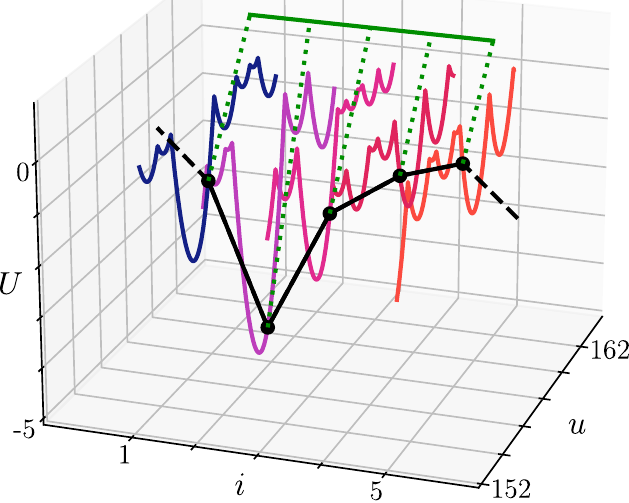}
    \caption{
        Schematic of our model with equation of motion in \cref{eq:motion} in dimension $d = 1$ for $L = 5$ particles.
        The potential energy landscape $U_i$ of each particle $i$ is a sequence of parabolic wells with uniform curvature $\mu = 1$ and random widths (that are quenched).
        In this example, the landscape of each particle is drawn in a different colour.
        The particle positions $u_i$ are indicated by the black dots.
        Their elastic interactions are schematically illustrated by black lines (with dashed lines indicating periodicity).
        The driving frame is indicated by a green line.
        It is connected to each particle by a weak spring of stiffness $k_f$ illustrated using a dotted green line.
    }
    \label{fig:model}
\end{figure}

4) Finally, we add a small damping term with uniform viscosity $\eta$ to each particle.
Underdamped dynamics correspond to $\eta^2 < 4 \mu m$.
We use $\eta = 0.1$ ($\ll 4 \mu m \equiv 4$).
We study this model numerically by integrating \cref{eq:motion} using the velocity Verlet algorithm, which uses a discrete time step $\Delta t = 0.05$~\footnote{
    Such that $\Delta t \ll \omega^{-1}$, with $\omega^2 = \mu / m - (\eta / (2m))^2 \approx 1$.
}.
See \cref{fig:model} for a schematic of the model for $L = 5$ particles.

\paragraph{Disorder}

For each particle, the potential energy landscape $U_i$ is a sequence of parabolic wells with uniform curvature $\mu = 1$ and random widths (that are quenched).
The elastic force $f^p_i = \mu (u_i - u^{\min}_i)$, with $u^{\min}_i$ the position of the local minimum $u^{\min}_i = (u^s_i + u^{s + 1}_i) / 2$ with $u^s_i = \sum_{y = 1}^s u^y_i$ the position of the edge of the well that the particle is currently in.
The width of the well, $u^{s + 1}_i - u^s_i$, is drawn randomly from a Weibull distribution with shape parameter $k = 2$ \footnote{
    The cumulative distribution of the widths $w$ is $\Phi(w) = 1 - \exp(-(w / w_0)^k)$ with $k = 2$ and $w_0 = 2$.
    A small offset of $10^{-5}$ is added such that $w > 0$ (i.e.~$\Phi(w \leq 0) = 0$).
    Note that the average width is $\langle w \rangle \approx 2$, such that the typical yield force is approximately $1$ (as we use stiffness $\mu = 1$).
    Finally, the sequence of wells for each particle $i$ is offset by a random number, such that the local minima around $u_i = 0$ are not correlated.
}.
A crucial quantity is the force needed to fail locally in the forward direction, $x_i \equiv \mu (u^{s + 1}_i - u_i)$.

\paragraph{Quasi-static loading}

Quasi-static loading corresponds to driving the frame $\bar{u}$ at a zero rate $\partial_t \bar{u} = 0$.
In practice, we increase $\bar{u}$ by a small amount and then follow the dynamics until the energy is minimized.
This sequence is then repeated until we gathered sufficient statistics (discarding start-up effects).
We save significantly on computational time using an event-driven algorithm such that we skip periods with a complete absence of local yielding events, see \cref{sec:event} for details.
These ``events'' are the primary object of most of our analysis.
We drive using a weak spring of stiffness $k_f = 1 / L^2$, whose magnitude matches the stiffness of the modes of wavelength $L$, so as not to affect the roughness of the line on scales much smaller than $L$.
We note that in an experimental setup, $k_f$ is the compliance of the loading mechanism, and could thus be varied.

\paragraph{Triggering}

The quasi-static response provides us with events at different forces $f$, whereby the response is purely elastic in between events.
Since we are interested in the statistics of avalanches at fixed $f$, we additionally manually trigger avalanches at different $f$ to efficiently gather sufficient statistics.
Thereby, we move a randomly selected particle over the first barrier in the forward direction while keeping the position of the driving frame fixed, see \cref{sec:triggering} for details.

\section{Nucleation}

To test \cref{eq:ellc}, we first estimate $f_c$.
We then establish a scaling relation for $\nu$.
Finally, we test \cref{eq:ellc} using measurements of $\ell_c(f)$.

\paragraph{First system-spanning event}

\begin{figure}[htp]
    \centering
    \includegraphics[width=\linewidth]{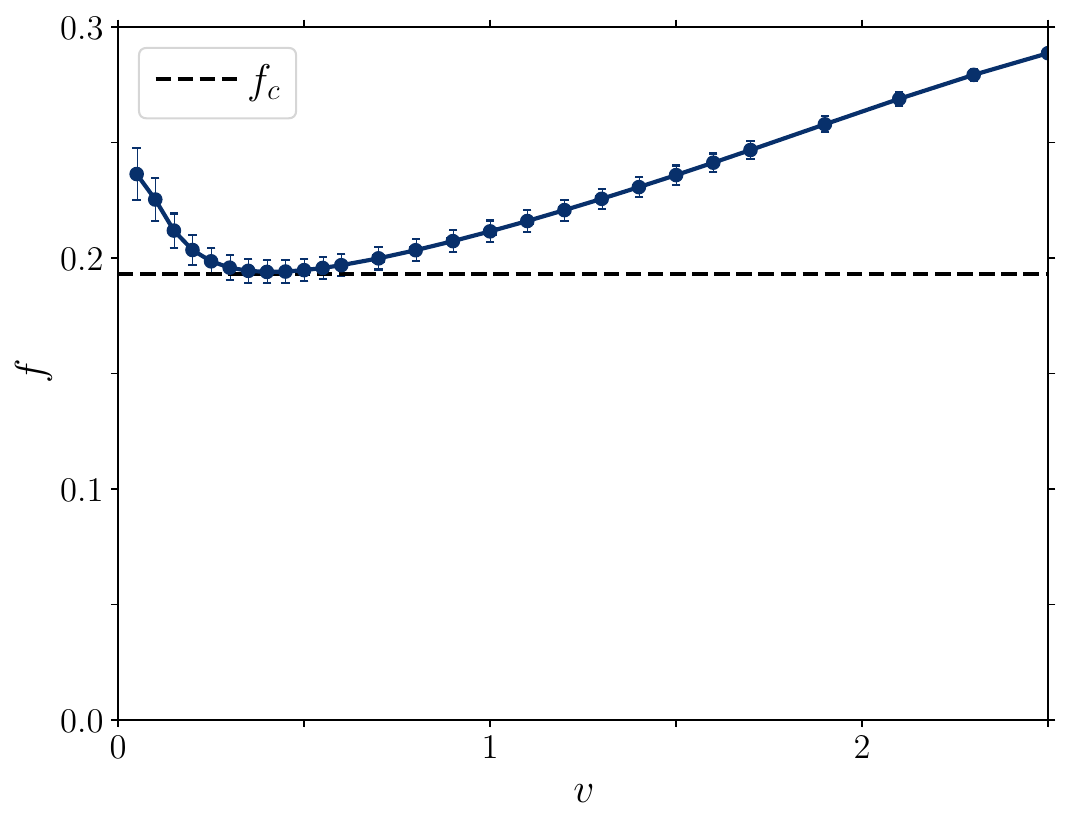}
    \caption{
        Measured flow curve, showing that the force $f$ at the driving frame is a non-monotonic function of the velocity $v$.
        The measurement of the unstable branch (where $\partial_v f < 0$) is possible by driving with a stiff spring.
        Also indicated: the critical force, $f_c$, defined the lowest force at which we find system-spanning events.
    }
    \label{fig:rheology}
\end{figure}

We first define $f_c$ as the lowest force at which we observe system-spanning events, if we manually trigger avalanches after a slip event during quasi-static loading.
This estimate, done in our largest system, is shown as a dashed line in \cref{fig:rheology}.

\paragraph{Flow curve}

In practice, the threshold force $f_c$ is predicted to be very close to the minimum of the flow curve $f_{\min}$ \cite{Brener2018,deGeus2023}.
We thus test the approximation $f_c \approx f_{\min}$.
To measure the latter, we need to measure the flow curve by driving the interface at a finite rate.
However, at small velocities, stick-slip occurs.
A classical approach to stabilize the velocity-weakening branch of the flow curve is to drive the system with a stiff spring \cite{Rice1983}, see also \cref{sec:kc}.
We use $k_f = 0.1$ (as a reference, we note that we recorded the quasi-static response above for $k_f = 1 / L^2 \approx 1.5 \times 10^{-8}$).
We plot the flow curve in \cref{fig:rheology} and find its minimum, confirming that $f_c \approx f_{\min}$.
In the \cref{sec:rheology}, we show that our measurement of $f_{\min}$ is robust with changing $k_f$, as long as the value of $k_f$ and the driving velocity are large enough (since stick-slip occurs otherwise).
In that figure, it is also shown that relaxation experiments give the same estimate for $f_{\min}$.

\paragraph{Scaling relation for $\nu$}

To measure the exponent $\nu$, we use the scaling relation relating it to the exponent $\zeta$ characterizing the roughness of the interface (i.e.~$u\sim \ell^\zeta$):
\begin{equation}
    \label{eq:nu}
    \nu = 1 / (2 - \zeta).
\end{equation}
This is a classical result of the depinning transition, which can be derived based on certain symmetries of the problem.
Intuitively, it is simply the statement that on the correlation length $\xi\sim (F-F_c)^{-\nu}$, the fluctuations of elastic force $\delta F\sim \xi^{\zeta-2}$ are of the order of the distance to threshold $F - F_c$.
This relation was observed to hold in the presence of inertia \cite{deGeus2023}.

\paragraph{Scaling relation for $\zeta$}

\begin{figure}[htp]
    \centering
    \includegraphics[width=\linewidth]{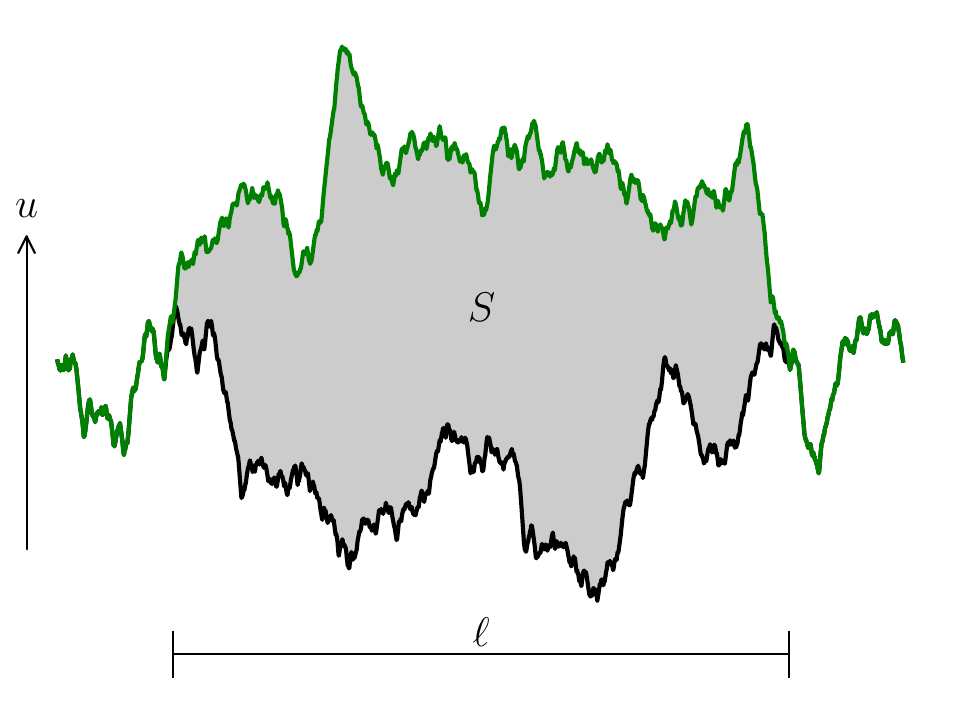}
    \caption{
        Position of the interface $u$ before (black) and after (green) an avalanche triggered at $f \approx f_c$.
        The total number of fails, $S$, is a proxy for the total displacement of the interface (in gray).
        The number of particles that fail at least once, $\ell$ (in $d = 1$), is a proxy for the linear extension of the avalanche (annotated).
    }
    \label{fig:geometry}
\end{figure}

\begin{figure}[htp]
    \centering
    \includegraphics[width=\linewidth]{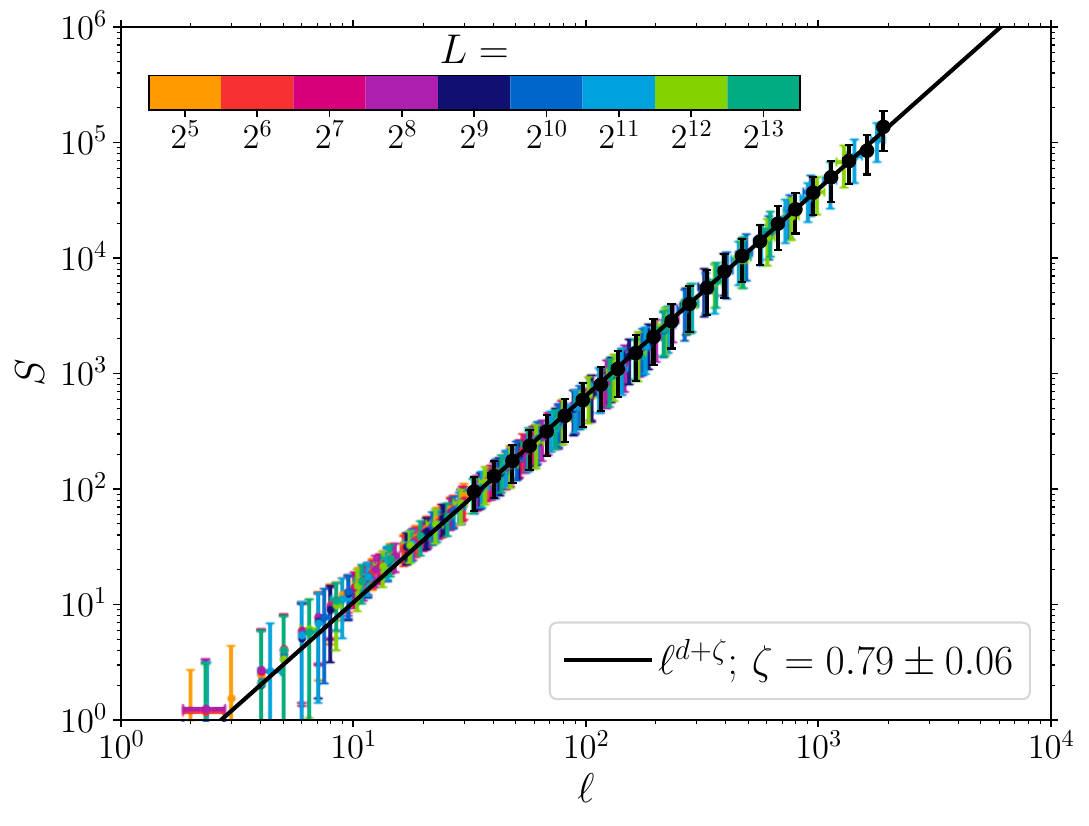}
    \caption{
        Measurement of the fractal dimension at $f \approx f_c$: the size $S$ of avalanches as a function of their extension $\ell$, and a fit of the scaling relation of \cref{eq:fractal} $S \sim \ell^{d + \zeta}$ as a solid black line (fit on the binned data for all avalanches for all considered system sizes, shown using black markers).
        The coloured markers correspond to a binning of individual (manually triggered) avalanches for different $L$.
        Note that we only show avalanches and remove system-spanning events (which at $f_c$ is only a small fraction of the events).
    }
    \label{fig:S-ell}
\end{figure}

To estimate the roughness exponent $\zeta$, we use the fact that it characterizes how the accumulated total displacement $S$ of an avalanche depends on its spatial extent $\ell$, as illustrated in \cref{fig:geometry}:
\begin{equation}
    \label{eq:fractal}
    S \sim \int \delta u \, d x^d \sim \ell^{d + \zeta},
\end{equation}
where we integrate over $d$ dimensions \footnote{
    Note that this exponent is commonly named the fractal dimension $d_f$ (i.e.\ $S \sim \ell^{d_f}$ with $d_f = d + \zeta$).
}.
Practically, we define the size $S$ as the total number of fails, and the extension $\ell$ as the number of particles that fail at least once (since $d = 1$).
Our results are shown in \cref{fig:S-ell}, indicating that $\zeta\approx 0.8$, which corresponds to $\nu \approx 0.83$.

\paragraph{Nucleation radius}

\begin{figure}[htp]
    \centering
    \includegraphics[width=\linewidth]{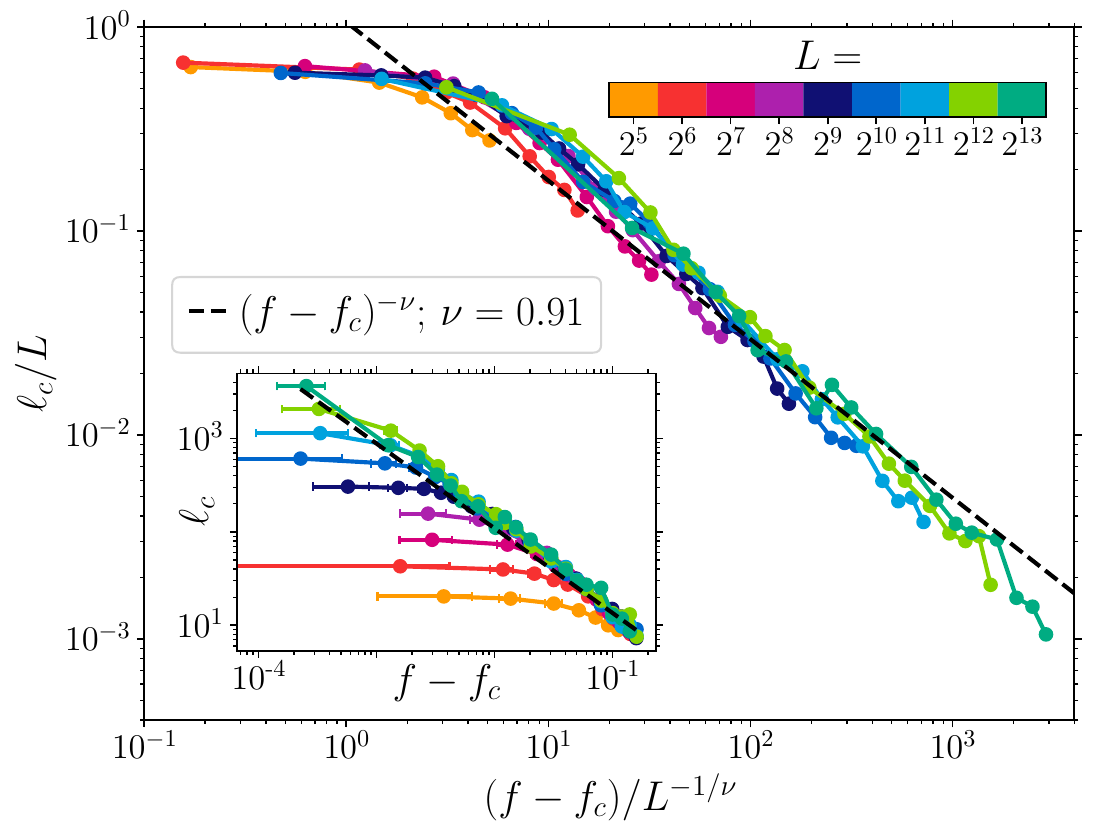}
    \caption{
        Characterization of the nucleation length $\ell_c$ beyond which avalanches are unstable, as a function of force $f$.
        We define $\ell_c$ as the upper cut-off of the distribution of avalanche's linear extension: $\ell_c = \langle \ell^3 \rangle / \langle \ell^2 \rangle$ (with $\langle \ldots \rangle$ the average over events that are not system-spanning, i.e.~$\ell < L$).
        The best collapse is obtained for $\nu \approx 0.9$ as shown in the main panel.
        The inset shows the uncollapsed data consistent with scaling $\ell_c \sim (f - f_c)^{-\nu}$.
    }
    \label{fig:ellc}
\end{figure}

We measure the nucleation radius $\ell_c(f)$ using the distribution of events triggered at different $f$.
We consider only the statistics of avalanches, i.e.~we discard the mode in the distribution corresponding to system-spanning events with $\ell = L$.
$\ell_c(f)$ corresponds to the cut-off of the distribution of avalanche extension, which we estimate as is often done as $\ell_c = \langle \ell^3 \rangle / \langle \ell^2 \rangle$, where the average is made on all avalanches.
Results are shown in the inset of \cref{fig:ellc}.

To test \cref{eq:ellc}, we employ a finite size collapse in \cref{fig:ellc} whereby we find $\nu \simeq 0.9$, in rather good agreement with the measurement of the avalanche fractal exponent in \cref{fig:S-ell}, which led to $\nu\approx 0.83$, using the scaling relation \cref{eq:nu}.
Supporting measurements based on a different protocol to extract $\ell_c$ are presented in \cref{sec:ellc}.
Thus, the fact that nucleation is controlled by a critical point, despite that the flowing regime displays a first order transition, appears to hold in short-range depinning as well.

\section{Effect of system size on the magnitude of the hysteresis cycle}

From our viewpoint, there exists a tiny hysteresis in the thermodynamic limit $f_c - f_{\min}$, which, however, is too small for us to resolve.
Yet, the observed hysteresis (the amplitude of the stick-slip cycle) is actually very large in our numerics, which we now explain.

\paragraph{Theoretical arguments}

We obtain a prediction for the amplitude of the stick-slip cycle $\Delta F(L)$ if we combine (i) the rate at which avalanches are nucleated, and (ii) the statistics of avalanches, with the above picture of nucleation.

\paragraph{(i) Armouring of the interface}

Significant mechanical noise is generated during a system spanning event.
This has the consequence that it is unlikely to find regions that are about to fail after such an event (as they would have failed during the event).
This is reflected in the additional force $x$ needed to trigger avalanches locally.
It was argued that such effect should lead to a density that follows \cite{deGeus2019, ElSergany2023} $P(x) \sim x^{\theta'}$.
As a result, the number of avalanches, $n_a$, triggered as the force is increased from $f_{\min}$ to $f_{\min}+ \Delta f \approx f_{c}+ \Delta f $ follows:
\begin{equation}
    \label{eq:na}
    n_a \sim L\int_0^{\Delta f} P(x) d x \sim L(\Delta f)^{\theta' + 1}.
\end{equation}

\begin{figure}[htp]
    \centering
    \includegraphics[width=\linewidth]{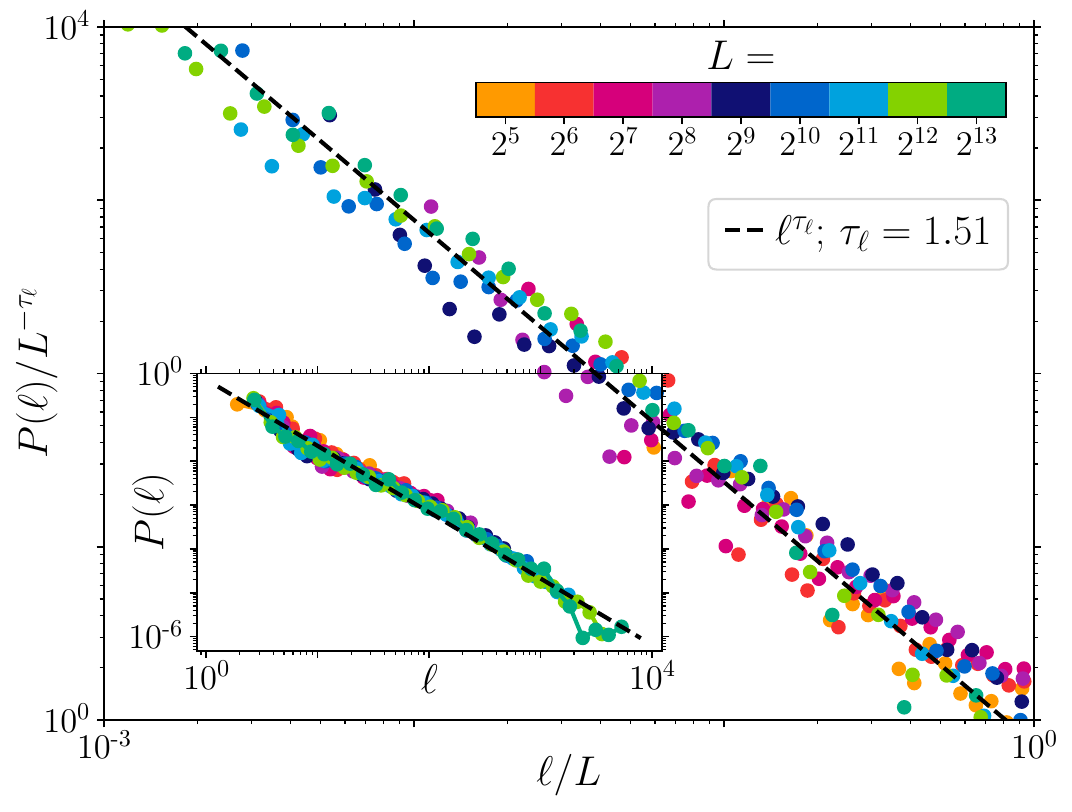}
    \caption{
        Distribution of avalanche extension, $P(\ell)$, at $f \approx f_c$.
        The exponent $\tau_{\ell}$ is shown using a dashed black line.
        The inset shows $P(\ell)$ while the main panel shows the finite size collapse with that value of $\tau_{\ell}$.
    }
    \label{fig:tau_ell}
\end{figure}

\paragraph{(ii) Avalanche statistics}

Near $f_c$, the distribution of avalanche extension is power law: $P(\ell)\sim \ell^{-\tau_\ell}$.
To measure $\tau_{\ell}$, we measure the distribution of the avalanche extension, $P(\ell)$, at $f \approx f_c$ as shown in \cref{fig:tau_ell}.
We obtain $\tau_{\ell}\approx 1.5$.

\paragraph{Maximal extension of avalanches and nucleation}

\begin{figure}[htp]
    \centering
    \includegraphics[width=\linewidth]{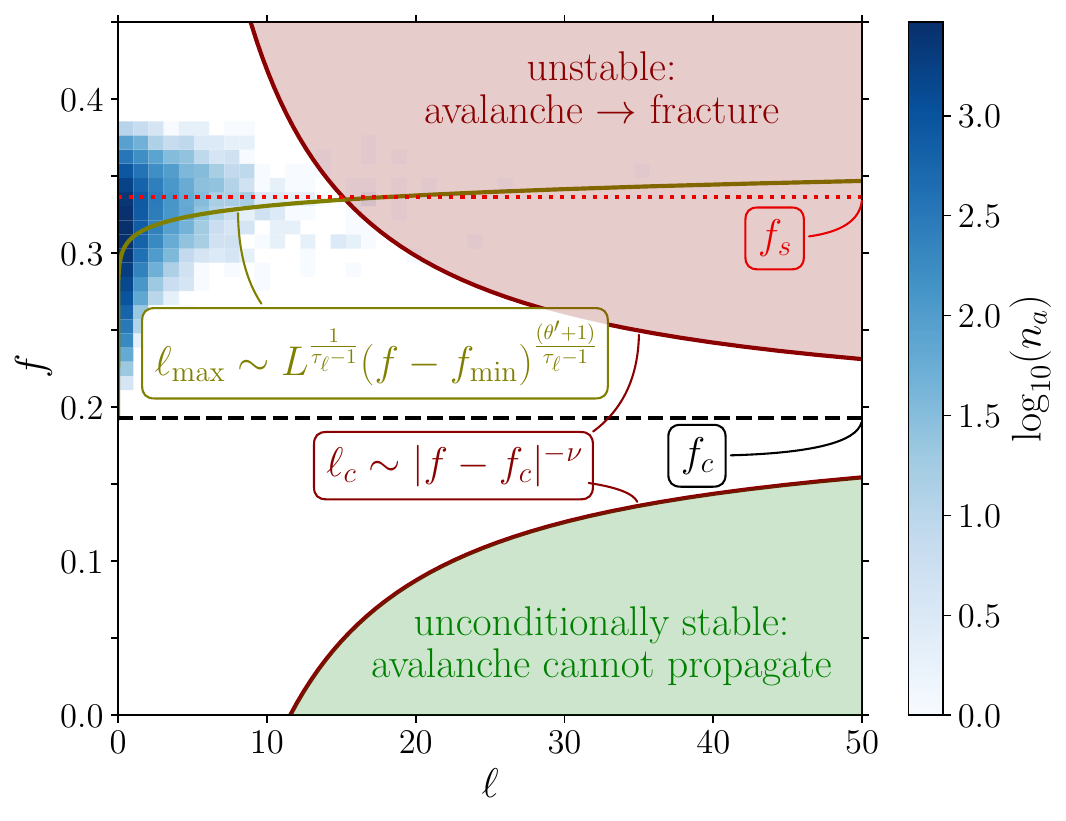}
    \caption{
        Spontaneous nucleation of slip during the stick-slip cycle in a finite system.
        Once stopped at $f_{\min}$ after a slip event, the system is ``armoured'' such that there are very few regions close to yielding.
        Due to this effect, the scale $\ell_{\max}$ of the largest avalanche grows very slowly with increasing force (gold curve).
        When this curve hits the nucleation size $\ell_c$ (red line), nucleation occurs and a system spanning event is triggered.
        In blue: the number of avalanches with a certain extension for avalanches triggered naturally during quasi-static loading, as indicated by the colour bar.
        The lines corresponding to the predictions of $\ell_{\max}$ (gold) and $\ell_c$ (red) are based on measured exponents, with arbitrary prefactors.
        The green region is indicated for completeness, as the region where avalanches cannot propagate, see \cite{deGeus2022}.
    }
    \label{fig:picture}
\end{figure}

Given $n_a$ avalanches, if $\ell_{\max}$ denotes the typical extension of the largest avalanche, extreme value statistics \cite{Bouchaud1990} implies that:
\begin{equation}
    n_a \int_{\ell_{\max}}^{\infty} P(\ell) d \ell \sim 1
\end{equation}
leading to:
\begin{equation}
    \label{eq:lmax}
    \ell_{\max} \sim n_a^{-1 / (1 - \tau_\ell)} \,.
\end{equation}
Nucleation will occur when $\ell_{\max}$ reaches the nucleation length $\ell_c \sim \Delta f^{-\nu}$.
This situation corresponds to the intersection of the gold and red lines in \cref{fig:picture}, which illustrates the present argument.

\paragraph{Stick-slip amplitude}

Using \cref{eq:na,eq:lmax} leads to the main result:
\detail{
    \begin{equation}
        L^{\frac{1}{\tau_\ell - 1}} (f_s - f_c)^{\frac{\theta' + 1}{\tau_\ell - 1}} \sim (f_s - f_c)^{-\nu}
    \end{equation}
    \begin{equation}
        L(f_s - f_c)^{\theta' + 1} \sim (f_s - f_c)^{\nu(1 - \tau_\ell)}
    \end{equation}
    \begin{equation}
        (f_s - f_c)^{\theta' + 1 - \nu(1 - \tau_\ell)} \sim L^{-1}
    \end{equation}
}
\begin{equation}
    \label{eq:stick-slip}
    \Delta f\sim L^{-1 / (\theta' + 1 - \nu(1 - \tau_\ell))}.
\end{equation}

\paragraph{Empirical tests}

\begin{figure}[htp]
    \centering
    \includegraphics[width=\linewidth]{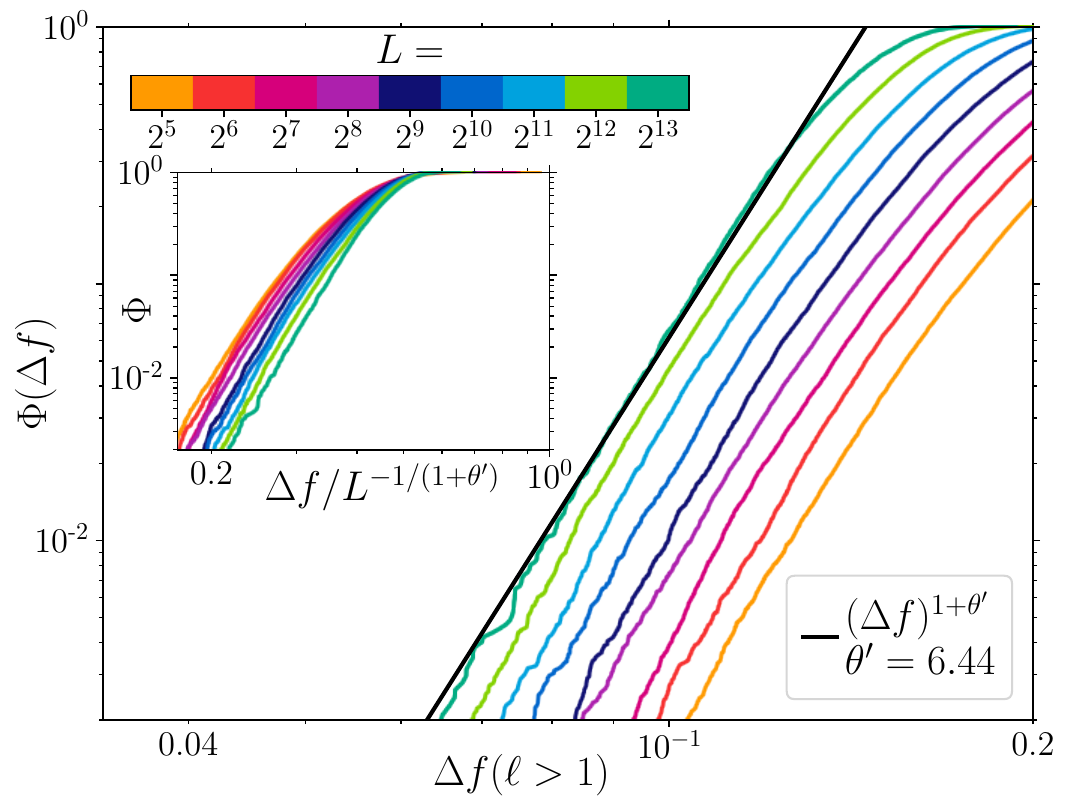}
    \caption{
        Cumulative number of avalanches, $\Phi$, upon increasing the driving force by $\Delta f$ after a system-spanning event, from which the exponent
        $\theta'$ is obtained.
        The inset shows how the data collapses with the finite size scaling predicted in \cref{eq:na}.
    }
    \label{fig:theta_prime}
\end{figure}

To test \cref{eq:stick-slip}, we first measure $\theta' \simeq 6.4$ in \cref{fig:theta_prime} by comparing the total number of collective events, events involving at least two rearrangements, upon increasing the driving force by $\Delta f$ after a system-spanning event \footnote{
    We have found, for the case of long-range elasticity~\cite{deGeus2019}, that the value of $\theta'$, representing the actual density of avalanches that are trigged upon increasing the force after system-spanning events, does not correspond to the exponent characterizing the density of small local distances (in force) to the yielding threshold.
    We found this scenario to be present also in this case involving short-range elasticity, see \cref{sec:stability} and \cite{deGeus2019} (in particular its supplementary information) for a discussion.
}.

Finally, we measure the magnitude of the hysteresis cycle $\Delta f(L)= f_s - f_{\min}$, where $f_s$ is the average force right before a spanning event (in red) and $f_{\min}$ the average force right after it right (in blue).
The scaling prediction of \cref{eq:stick-slip} is tested in inset, and shows an excellent agreement with our predictions.

\begin{figure}[htp]
    \centering
    \includegraphics[width=\linewidth]{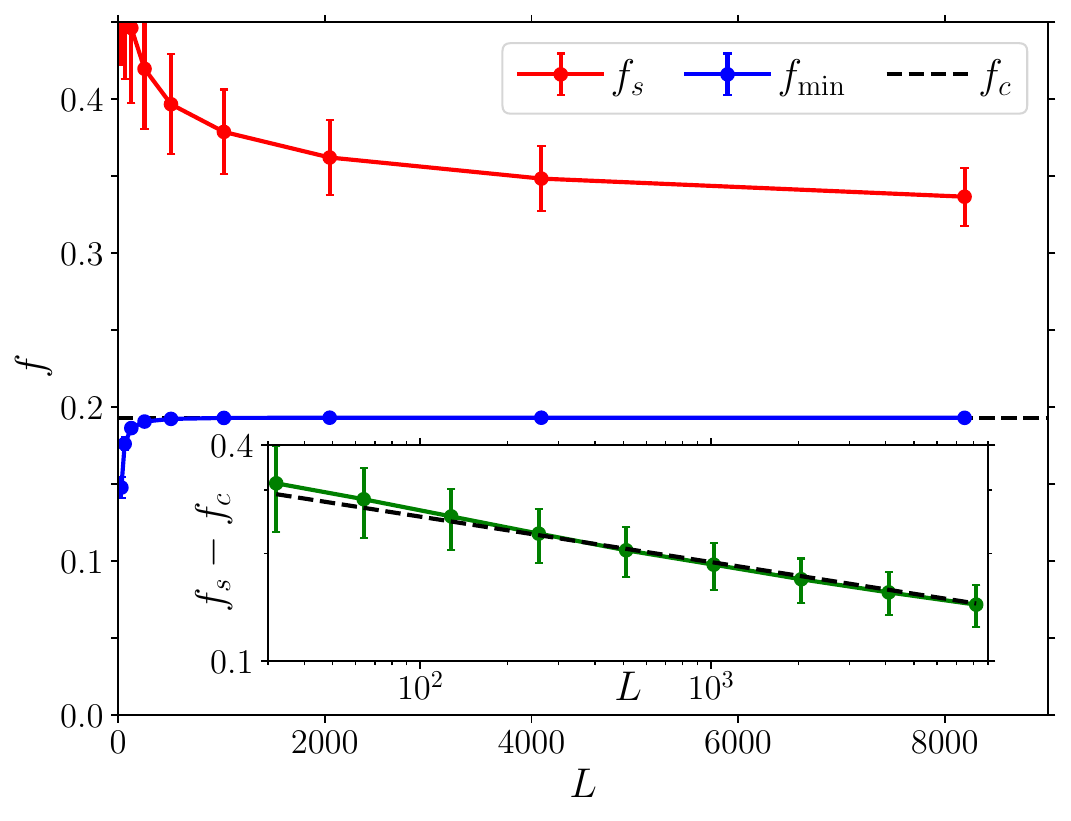}
    \caption{
        Main panel: Finite size scaling of the force just before system-spanning events ($f_s$ in red) and just after system-spanning events ($f_{\min}$ in blue).
        The markers correspond to the means and the error bars to the standard deviation.
        Inset: finite size scaling of the stick-slip amplitude $f_s - f_c$ with $f_c$ the smallest force at which a system-spanning event was found to occur in any of our finite systems (also shown in the main panel using a dashed black line).
        The dotted line corresponds to our prediction in \cref{eq:stick-slip}, with exponents taken from \cref{fig:tau_ell,fig:theta_prime,fig:ellc}.
    }
    \label{fig:finite-size}
\end{figure}

\section{Conclusion}

We have tested a theory for the short-range depinning transition when velocity weakening is present, as can occur e.g.~for weak damping when inertia plays a large role.
This theory builds on the continuous rate-and-state framework to describe homogenous frictional interfaces, and treat disorder perturbatively \cite{deGeus2022}.
We have confirmed in the case of a pinned elastic line that in the presence of velocity weakening, the depinning transition has a surprising mixed behaviour.
Specifically, the flowing phase has properties of a first order phase transition: the flow curve is non-monotonic, leading to a finite amount of hysteresis.
When nucleation occurs, it generates flow at a finite speed.
Yet, the avalanche-type response in the solid phase is controlled by a continuous critical point, predicted to be just above the minimum of the flow curve.
Scale free avalanches are present, which control the nucleation process.
The latter leads to immense finite size effects on the magnitude of hysteresis.
This phenomenology is reminiscent of faults, paradoxically known to display power-law distributed avalanches as well as fault-spanning earthquakes that can be suddenly triggered, reminiscent of a nucleation process.

Note that the present model could be used to study larger damping, for which velocity-weakening may disappear and \ref{item:scenario:b} discussed in the introduction could apply.
It would also allow one to study how exponents cross-over between an overdamped and underdamped regime \cite{Salerno2012,Salerno2013,Ertas1992}.

Finally, another interesting question concerns the applicability of the proposed scenario to other systems.
A bimodal distribution of avalanches is also observed in amorphous solids under stress when inertia is important~\cite{Karimi2017}.
A pseudo-gap in the distribution of local yield stress must be present in these systems even in the overdamped limit \cite{Lin2014}, but it is enhanced by inertia that increases its associated exponent $\theta$~\cite{Karimi2017}.
It remains to be seen if the minimum of the flow curve controls the nucleation of rupture in that case as well.

\section*{Acknowledgements}

Damien Ribi\'{e}re is thanked for derivations concerning the stabilization of the velocity weakening instability by a stiff driving spring.
Alberto Rosso is thanked for discussions.
T.G.\ acknowledges support from the Swiss National Science Foundation (SNSF) by the SNSF Ambizione Grant PZ00P2{\_}185843.
M.W.\ acknowledges support from the Simons Foundation Grant (No.~454953 Matthieu Wyart).

\appendix

\renewcommand\theequation{A\arabic{equation}}
\setcounter{equation}{0}

\renewcommand\thefigure{A\arabic{figure}}
\setcounter{figure}{0}

\renewcommand\thetable{A\arabic{table}}
\setcounter{table}{0}

\section{Model}

\subsection{Interactions}
\label{sec:interactions}

Following \cite{Rosso2001}, we consider elastic interactions with a potential energy density that we expand to the fourth order:
\begin{equation}
    E(\bm{x}) = \tfrac{1}{2} k_2 \big(\varepsilon(\bm{x}) \big)^2 + \tfrac{1}{12} k_4 \big(\varepsilon(\bm{x}) \big)^4,
\end{equation}
with $\bm{x}$ the spatial coordinate along the interface, $k_2$ and $k_4$ stiffnesses, and the gradient of the position $\varepsilon = || \partial_{\bm{x}} u ||$.
The resulting force $f^n(\bm{x}) = \mathrm{div}\, \partial_{\bm{\varepsilon}} E(\bm{x})$.
In $d = 1$ the geometry corresponds to an elastic line, for which
\begin{equation}
    f^n(x)
    = \left. \frac{\partial^2 u}{\partial x^2} \right|_{x}
    \left( k_2 + k_4 \left(\left. \frac{\partial u}{\partial x} \right|_{x} \right)^2 \right),
\end{equation}
which we discretize as~\cite{Rosso2002a}:
\begin{equation}
    \begin{split}
        f_i^n &= k_2 \big[ u_{i - 1} - 2 u_i + u_{i + 1} \big] \\
        &+ k_4 \big[ (u_{i + 1} - u_i)^3 + (u_i - u_{i - 1})^3 \big].
    \end{split}
\end{equation}
\detail{
    Alternative:
    \begin{equation}
        \begin{split}
            f_i^n &= k_2 \big[ u_{i - 1} - 2 u_i + u_{i + 1} \big] \\
            &+ k_4 \big[ (u_{i + 1} - u_i)^3 + (u_i - u_{i - 1})^3 \big]
        \end{split}
    \end{equation}
}
We use $k_2 = k_4 = 1$~\footnote{
    For overdamped dynamics, this model has as known limitation in $d = 1$ that linear elasticity ($k_4 = 0$) results in an unphysical roughness $\zeta_\infty > 1$, while $\zeta_\infty = 0.63$ if $k_4 > 0$~\cite{Rosso2001,Rosso2002a}.
    This is unimportant for our scaling prediction.
    However, we stick to the physical case and choose $k_2 = k_4 = 1$ in $d = 1$.
}.
Note that we have expressed $x$ in units of the `discretization' $h = 1$, such that the length of the line is $L$.
Finally, we assume periodic boundary conditions such that $u_1 = u_L$ \footnote{
    In fact, $u_{L + i} = u_i$ for any $i$.
}.

\subsection{Event-driven quasi-static loading}
\label{sec:event}

Since the system is linear as long a no particle fails, we know the particle positions $u_i$ satisfying the mechanical equilibrium for any driving frame position $\bar{u}$ that matches the condition that all particles stay in their current potential well.
This allows the following protocol.
In alternation, we set $\bar{u}(t + \Delta t) = \bar{u}(t) + \Delta u (1 + \mu / k_f)$ and $u_i(t + \Delta t) = u_i(t) + \Delta u$ for all $i$, as follows.
First, we perform a fully elastic step whereby the displacement $\Delta u$ is chosen such that the particle closest to failing in the direction of driving ends at the verge of failing.
In particular, $\Delta u = (\min (x_i) / \mu - \epsilon / 2)_+$ (with the Macaulay brackets $(\ldots)_+$ indicating that the term is zero if the argument is negative, and $\epsilon = 10^{-3}$).
Since, by definition, no particle fails during this step, the system stays in mechanical equilibrium.
Second, we apply a ``kick'' and set $\Delta u = \epsilon$ followed by energy minimization using the dynamics in \cref{eq:motion}.
Whatever happens during this step we refer to as an ``event'', and they are the primary object of most of our analysis.

We collect the following number of events during quasi-static loading, for a system with $L$ particles: $L = 2^5$: 357818, $L = 2^6$: 199964, $L = 2^7$: 199953, $L = 2^8$: 199934, $L = 2^9$: 84416, $L = 2^{10}$: 64224, $L = 2^{11}$: 72025, $L = 2^{12}$: 89594, $L = 2^{13}$: 32257.

\subsection{Triggering}
\label{sec:triggering}

For different $f$ and starting from each system-spanning event, we select the state with the highest $f_q < f$ acquired by normal quasi-static loading (hence the subscript $q$).
We elastically load that state up to $f$ (by advancing the driving frame by $\Delta \bar{u} = (f - f_q) / k_f$, and moving $u_i$ accordingly to maintain mechanical equilibrium, see \cref{sec:event}).
Then, while keeping the boundary condition $\bar{u}$ fixed, we move a particle over the first barrier in the forward direction such that $u_i = u^{s + 1}_i + \epsilon / 2$ for a randomly selected\footnote{
    For a fraction of particles, the particle's failure will not trigger an avalanche.
    To save on computational time, we rapidly assess, using entirely heuristic measures, if that is the case.
    If so, we select another particle.
    This does not result in any quantitative data about which particles are more likely to trigger an avalanche, since, without energy minimization, we cannot ensure if the particle truly did not trigger an avalanche.
    Precisely that energy minimization we chose to avoid by using heuristic rules, to save on computational time.
} $i$ and minimize energy.
In practice, starting from each system-spanning event, we load with different $\Delta f$ and then bin on $f$.

\section{Bimodal distribution}
\label{sec:criticality}

\begin{figure}[htp]
    \centering
    \includegraphics[width=\linewidth]{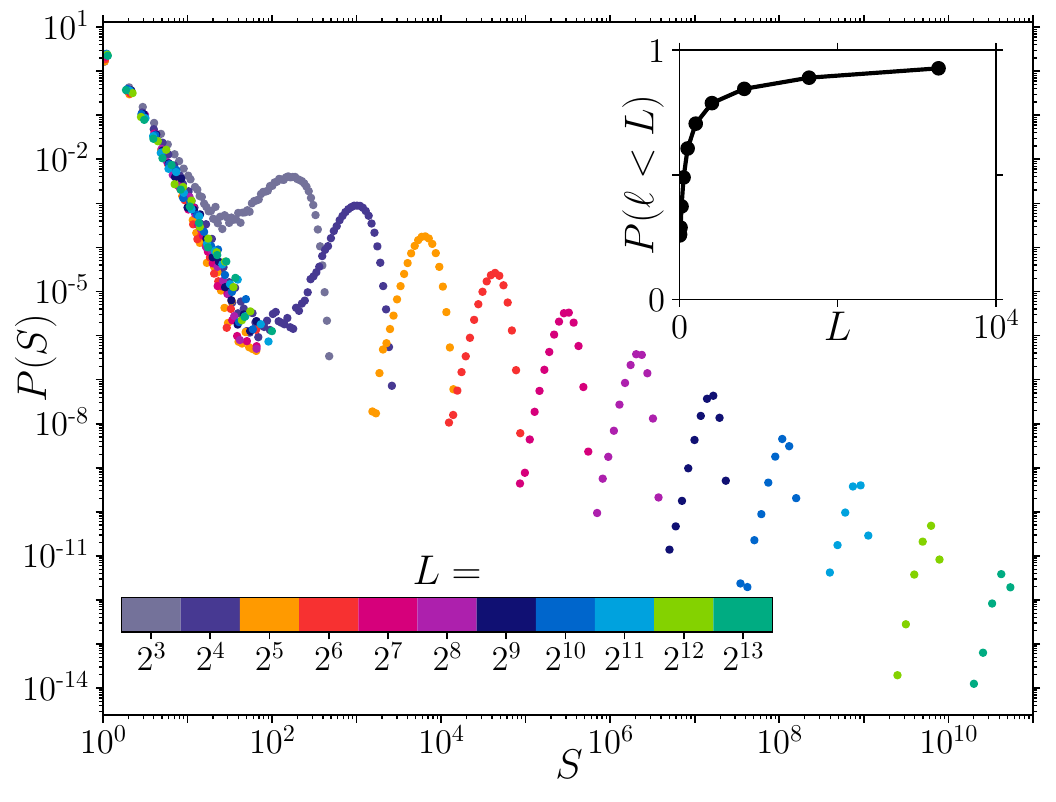}
    \caption{
        Distribution of the event sizes $S$ of any event (``avalanche'' or ``system-spanning'') during quasi-static loading, for different system sizes as indicated using different colours (see values of the system size $L$ in the colour bar).
        \emph{Avalanches}, cut-off at a small radius, constitute the majority of events (inset: fraction of events that are avalanches as a function of system size $L$).
        \emph{System-spanning}, or ``slip'', events span the entire system, and constitutive the right-hand side peak whose position increases with the system size $L$.
    }
    \label{fig:ps}
\end{figure}

During the stick-slip cycle, we observe two populations of collective events: avalanches and system-spanning events.
Avalanches are the largest fraction of these collective events.
These are events in which one or more, \emph{but not all}, particles fail at least once, such that $\ell < L$.
We quantify this observation in \cref{fig:ps}.
In the main panel, the distribution of the event sizes $P(S)$ for different system sizes $L$ shown using different colours.
Thereby, $S$ represents the total displacement of the line, and is defined as the total number of fails during an event.
We clearly observe two populations of events: \emph{avalanches} on the left and side, and \emph{system-spanning} events on the right-hand side.
Only for small systems, these two modes of the distribution merge into a single one, see e.g.~\cite{Prado1992}.

The system-spanning events are ``slip'' events during which the system accumulates macroscopic slip, while sub-extensive avalanches occur during the ``stick'' phase of the stick-slip cycle.
The fraction of events that are avalanches is shown in the inset as a function of system size $L$.
As observed, this fraction approaches one for large $L$.

\section{Stability of rate-and-state friction}
\label{sec:kc}

Consider a single particle with mass $m$ that is driven by a spring of stiffness $k_f$.
Furthermore, it is subjected to velocity weakening rate-and-state friction.
In such a model, the friction force
\begin{equation}
    f_p = f_0 + a \ln (v_i / v_0) + b \ln (v_0 \chi / D_c).
\end{equation}
Here, $f_0$, $a$, and $b$ are parameters with the units of force here; $v_0$ sets the units of velocity; $\chi$ is the age of the interface; and $D_c$ is the sliding distance needed to rejuvenate the interface.
We complement this law with a simple ageing law
\begin{equation}
    \dot{\chi} = 1 - v_i \chi / D_c.
\end{equation}
For this law, the steady state age, after rejuvenation has completed, corresponds to
\begin{equation}
    \label{eq:si:ageing}
    \chi_\infty = D_c / v_i.
\end{equation}
We stabilize this rheology by adding a viscous term $\eta v_i$.

In the steady state, the resulting rheology is non-monotonic, with a minimum at $v_c = (b - a) / \eta$.
The unstable branch, where $\partial_v f < 0$, corresponds to $v < v_c$.
The stable branch, where $\partial_v f > 0$, corresponds to $v > v_c$.

Supposing that the particle is in a steady state in mechanical equilibrium (such that the particle exactly follows the drive as $v_i = v$), force balance corresponds to
\begin{equation}
    \label{eq:si:force-balance}
    \underbrace{f_0 + a \ln \left(\frac{v}{v_0}\right) + b \ln \left(\frac{v_0 \chi_\infty}{D_c}\right)}_{\text{rate-and-state}} + \underbrace{\eta v}_{\text{heat}} = \underbrace{k_f (v t - u_i)}_{\text{drive}}.
\end{equation}
We will now consider the effect of a small perturbation $v + \delta v$, $\chi_\infty + \delta \chi$, and $u_i + \delta u$.
This results in
\detail{
    \begin{equation}
        \begin{split}
            f_0 &+ a \ln \left(\frac{v}{v_0}\right) + \eta v \\
            &+ \delta v \frac{\partial}{\partial v'} \left[ a \ln \left(\frac{v'}{v_0}\right) + \eta v' \right]_{v' = v} \\
            &+ b \ln \left(\frac{v_0 \chi_\infty}{D_c}\right) \\
            &+ \delta \chi \frac{\partial}{\partial \chi'} \left[ b \ln \left(\frac{v_0 \chi'}{D_c}\right) \right]_{\chi' = \chi_\infty} \\
            &= k_f (v t - u_i - \delta u).
        \end{split}
    \end{equation}
    \begin{equation}
        \begin{split}
            f_0 &+ a \ln \left(\frac{v}{v_0}\right) + \eta v + \delta v (a / v + \eta) \\
            &+ b \ln \left(\frac{v_0 \chi_\infty}{D_c}\right) + \delta \chi b / \chi_\infty \\
            &= k_f (v t - u_i - \delta u).
        \end{split}
    \end{equation}
    Using \cref{eq:si:ageing} and the identity in \cref{eq:si:force-balance} we find
}
\begin{equation}
    \label{eq:si:perturbation:rs}
    \delta v (a / v + \eta) + \delta \chi b / \chi_\infty = -k_f \delta u.
\end{equation}
Applying the same perturbation to the ageing law (still around the steady state) results in
\detail{
    \begin{equation}
        \delta \dot{\chi}
        = \delta v \left. \frac{\partial \dot{\chi}}{\partial v'} \right|_{v' = v} +
        \delta \chi \left. \frac{\partial \dot{\chi}}{\partial \chi'} \right|_{\chi' = \chi_\infty}
    \end{equation}
    or
    \begin{equation}
        \delta \dot{\chi} = - \delta v (\chi_\infty / D_c) - \delta \chi (v_i / D_c).
    \end{equation}
    or
}
\begin{equation}
    \delta \dot{\chi} = - \delta v / v - \delta \chi / \chi_\infty.
\end{equation}
Combining it with the time derivative of \cref{eq:si:perturbation:rs} results in
\begin{equation}
    \begin{cases}
        & \delta \dot{v} (a / v + \eta) + \delta \dot{\chi} b / \chi_\infty = -k_f \delta v \\
        & \delta \dot{\chi} = - \delta v / v - \delta \chi / \chi_\infty
    \end{cases}
\end{equation}
Let us make the ansatz $\delta v = v_0 e^{\lambda t}$ and $\delta \chi = \chi_0 e^{\lambda t}$ (with $\lambda$ a complex number) and search for solutions.
After some algebra, we find
\detail{
    \begin{equation}
        \begin{cases}
            & \lambda v_0 e^{\lambda t} (a / v + \eta) + \lambda \chi_0 e^{\lambda t} b / \chi_\infty = -k_f v_0 e^{\lambda t} \\
            & \lambda \chi_0 e^{\lambda t} = -v_0 e^{\lambda t} / v - \chi_0 e^{\lambda t} / \chi_\infty
        \end{cases}
    \end{equation}
    \begin{equation}
        \begin{cases}
            & \lambda v_0 (a / v + \eta) + \lambda \chi_0 b / \chi_\infty = -k_f v_0 \\
            & \lambda \chi_0 = -v_0 / v - \chi_0 / \chi_\infty
        \end{cases}
    \end{equation}
    \begin{equation}
        \begin{cases}
            & - \chi_0 / \chi_\infty = (v_0 / b) (a / v + \eta) + k_f v_0 / (b \lambda) \\
            & \lambda \chi_0 = -v_0 / v - \chi_0 / \chi_\infty
        \end{cases}
    \end{equation}
    \begin{equation}
        \begin{split}
            &- \lambda \chi_\infty v_0 / b (a / v + \eta) - \lambda \chi_\infty k_f v_0 / (b \lambda) \\
            &= -v_0 / v + (v_0 / b) (a / v + \eta) + k_f v_0 / (b \lambda)
        \end{split}
    \end{equation}
    \begin{equation}
        \begin{split}
            &- \lambda \chi_\infty v_0 (a / v + \eta) - \chi_\infty k_f v_0 \\
            &= -b v_0 / v + v_0 (a / v + \eta) + k_f v_0 / \lambda
        \end{split}
    \end{equation}
    \begin{equation}
        \begin{split}
            & \lambda^2 (a / v + \eta) + \lambda k_f - \lambda b / (v \chi_\infty) \\
            & + \lambda / \chi_\infty (a / v + \eta) + k_f / \chi_\infty = 0
        \end{split}
    \end{equation}
}
\begin{equation}
    \lambda^2 \left(\frac{a}{v} + \eta\right) + \lambda \left(\frac{a - b + \eta v}{D_c} + k_f \right) + \frac{k_f v}{D_c} = 0
\end{equation}
Now, the perturbation will grow exponentially in time if $\text{Re}(\lambda) > 0$, while they will decay exponentially if $\text{Re}(\lambda) < 0$.
The critical case therefore corresponds to $\text{Re}(\lambda_c) = 0$, such that $\lambda_c = i \text{Im}(\lambda)$.
This implies that $-\text{Im}(\lambda_c)(a / v + \eta) + k_f v / D_c = 0$ and $\text{Im}(\lambda_c) ((a - b + \eta v) / D_c + k_f) = 0$.
From the latter, it follows that the critical stiffness is
\begin{equation}
    k_f = (b - a - \eta v) / D_c \equiv k_c.
\end{equation}
If $k_f < k_c$, a perturbation will grow exponentially, and the system is unstable.
If $k_f > k_c$, a perturbation will decay exponentially, and the system is stable.
In this case, the particle will not show stick-slip even if driven at a rate $v < v_c$.

\begin{figure}[htp]
    \centering
    \includegraphics[width=0.7\linewidth]{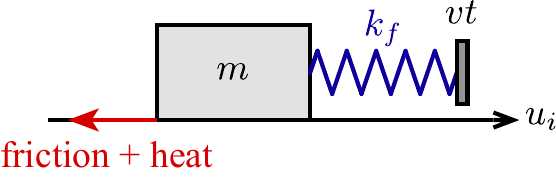}
    \caption{
        Simple model of a single particle subject to a rate-and-state friction force and heat radiation, and driving at a velocity $v$ by a weak spring of stiffness $k_f$.
    }
    \label{fig:particle-model}
\end{figure}

\section{Measuring \texorpdfstring{$f_{\min}$}{fmin}}
\label{sec:rheology}

\paragraph{Driven interface stabilised by a stiff spring}

\begin{figure}[htp]
    \centering
    \includegraphics[width=\linewidth]{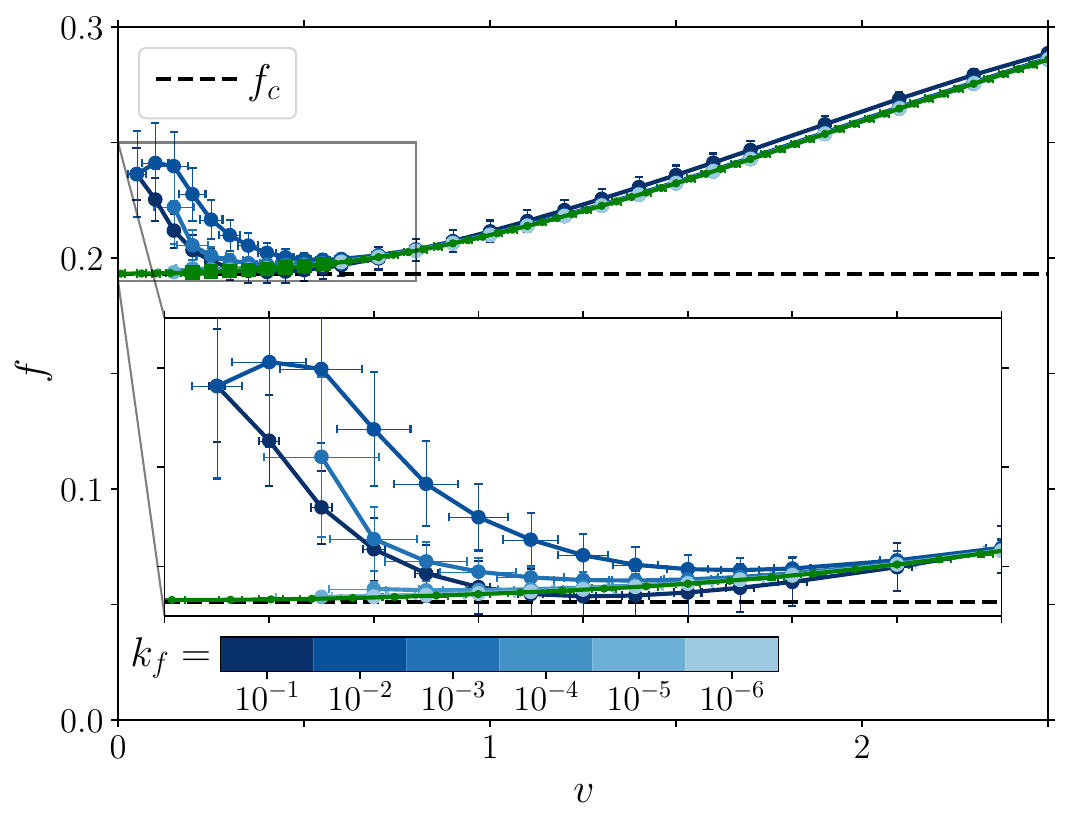}
    \caption{
        For the largest system.
        Green curve: relaxation measurement: the force and velocity are measured during a system-spanning event.
        The shown data is the average force per bin of velocity.
        The large square green markers correspond to measurements of the force if the system is driven at a constant, finite rate.
        Measurements in which the system intermittently stops are discarded.
        The same type of measurement is shown for different driving stiffnesses in blue (see colour bar).
    }
    \label{fig:relaxation}
\end{figure}

We measure the rheology of our model by driving the system at a finite rate $v$ \footnote{
    At every time step we move the driving frame $\bar{u}(t + \Delta t) = \bar{u}(t) + v \Delta t$.
} and measuring the average force experienced by the driving frame $f$.
We consider only the steady state response.
Any transient start-up effects or intermittent response is discarded.
We consider our largest system $L = 2^{13}$ and measure the rheology for different stiffnesses $k_f$ (as a reference, we note that we recorded the quasi-static response above for $k_f = 1 / L^2 \approx 1.5 \times 10^{-8}$).
The results are shown in \cref{fig:relaxation}.
The different blue curves correspond to different stiffnesses $k_f$ (see colour bar).
The non-monotonic rheology is evidenced when driving with a stiff spring.
However, the rheology is an interplay between the response of the interface and the driving spring, as clearly observed.
We do not understand the details of this interplay, and from \cref{fig:relaxation} it seems non-trivial, as the order of the curves is not monotonic in $k_f$.
Objectively extracting $f_{\min}$ (and $v_{\min}$) from this data is therefore currently not possible, and future work is urged.

\paragraph{Relaxation}

We propose an alternative protocol to measure the rheology.
We measure the relaxation during a system-spanning slip event during quasi-static loading (again using the low, reference, $k_f$).
In particular, we measure the force experienced by the driving frame, $f$, as a function of the average velocity of the interface, $v$.
The result is shown in \cref{fig:relaxation} using a green curve.
This protocol cannot acquire the unstable branch of the rheology, but it does allow measuring the stable branch.
It is tempting to speculate that it does so until $v = v_c$.
If that is the case, we find a good agreement between the minimum of the rheology, $f_{\min}$, and the definition of $f_c$ that we used above.

\section{Alternative measurement of \texorpdfstring{$\ell_c$}{lc}}
\label{sec:ellc}

\begin{figure}[htp]
    \centering
    \includegraphics[width=\linewidth]{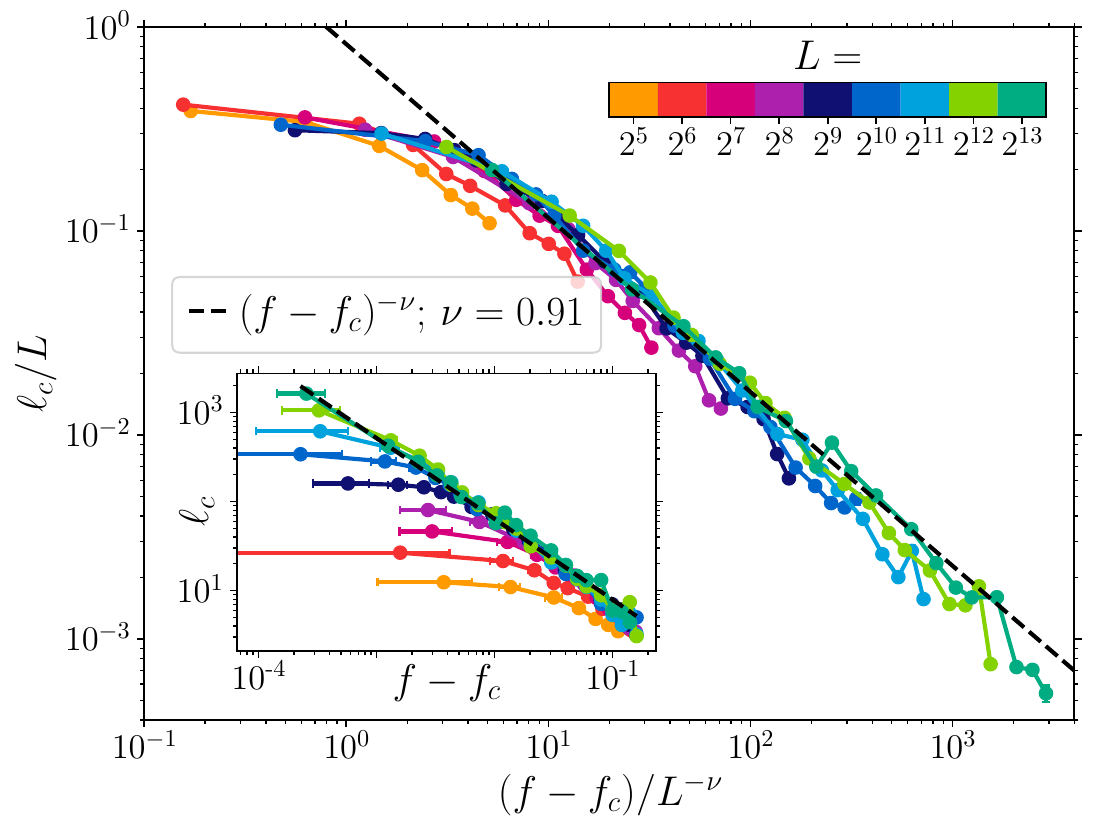}
    \caption{
        Complementary to \cref{fig:ellc} we measure $\ell_c$ by fitting an exponential cut-off to the cumulative distribution of the linear extension of avalanches (see text), and verify that our data is consistent with the prediction in \cref{eq:ellc,eq:nu}.
    }
    \label{fig:ellc_alternative}
\end{figure}

In the main text, we defined $\ell_c = \langle \ell^3 \rangle / \langle \ell^2 \rangle$.
Here we use an alternative protocol where we fit an exponentially cut-off power law to the cumulative distribution $\Phi(\ell) \sim \ell^{1 - \tau_\ell} \exp(-\ell / \ell_c)$.
The result in \cref{fig:ellc_alternative} is indeed also consistent with our prediction in \cref{eq:ellc,eq:nu}.

\section{Armouring}
\label{sec:stability}

\paragraph{Distribution of local yield stress $P(x)$}

\begin{figure}[htp]
    \centering
    \includegraphics[width=\linewidth]{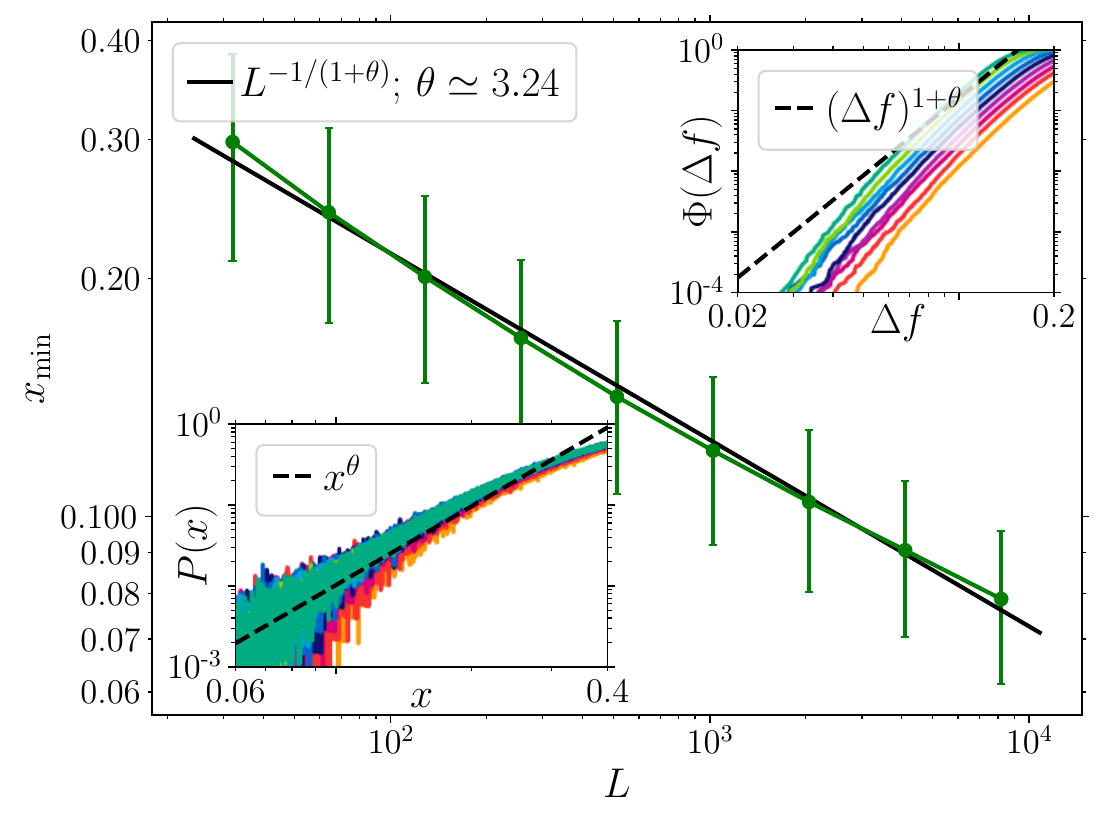}
    \caption{
        Characterization of (force) barriers after system-spanning events.
        The force needed to trigger failure locally $x_i \equiv \mu (u^Y_i - u_i)$ with $u^Y_i$ the position of the edge of the local potential in the direction of triggering.
        Plotted is $x_{\min}$ as a function of the number of particles $L$, which allows fitting the exponent $\theta$ of $P(x) \sim x^\theta$ as indicated (see also main text).
        $x_{\min}$ is taken as the mean of the minimum of $x$ after $L$ system-spanning events, see \cref{sec:event}.
        Lower-left inset: $P(x)$ for different system sizes.
        Upper-right inset: cumulative number of events upon increasing the driving force by $\Delta f$ after a system-spanning event.
        In both cases, the black line corresponds to the $\theta$ fitted in the main plot.
    }
    \label{fig:stability}
\end{figure}

The distribution of local yield stress after system-spanning events contains information about the number of plastic events that can be triggered upon increasing the load.
We measure $x_i$ as the additional force needed to fail a particle $i$ in the forward direction.
The density $P(x)$ is consistent with a pseudo-gap distribution
\begin{equation}
    \label{eq:px}
    P(x) \sim x^\theta,
\end{equation}
see lower-left inset in \cref{fig:stability}.
However, extracting $\theta$ from the noisy data is difficult.
The proper measurement of the exponent is therefore to perform extreme value statistics using finite size scaling.
Once again, we employ the result of \cite{Bouchaud1990} to argue that
\begin{equation}
    L\int_0^{x_{\min}} P(x) d x \sim 1,
\end{equation}
(with $L$ the number of particles).
Assuming the pseudo-gap distribution of \cref{eq:px}, we find that $x_{\min} \sim L^{-1 / (1 + \theta)}$.
Our data is consistent with this result with $\theta = 3.2$ as shown in the main panel of \cref{fig:stability}.
This result is not far from our earlier result of $\theta = 3.7$~\cite{deGeus2019} in our model based on long-range elasticity.
Furthermore, this value is of the same order as a prediction based on a single particle model \cite{ElSergany2023}, see~\cite{Purrello2020} for other results in a single particle model.

\paragraph{Only a vanishing fraction of plastic events can trigger avalanches}

Our main result depends on $\theta'$, which is based on the actual number of avalanches (in which at least two particles fail) upon increasing the load by $\Delta f$ after a system-spanning event.
Interestingly, we find that $\theta < \theta'$, cf.~\cref{fig:theta_prime,fig:stability}.
We thus find, as for long-range elasticity \cite{deGeus2019}, that not all failures trigger an avalanche.
As we showed in the supplementary material of \cite{deGeus2019}, we find non-trivial spatial correlations between barriers of particles after system-spanning events, likely corresponding to the different between $\theta$ and $\theta'$, however, in a way that is not yet fully understood.

\bibliographystyle{unsrtnat}
\bibliography{library}

\end{document}